# Molecular dynamics simulation of multivalent ion mediated DNA attraction

Liang Dai,[1] Yuguang Mu,[2] Lars Nordenskiöld,[2] and Johan R. C. van der Maarel[1]

[1]*National University of Singapore, Department of Physics, 2 Science Drive 3, Singapore 117542*
[2]*Nanyang Technological University, School of Biological Sciences, 60 Nanyang Drive, Singapore, 63755*
(January 8, 2008)

All atom molecular dynamics simulations with explicit water were done to study the interaction between two parallel double-stranded DNA molecules in the presence of the multivalent counterions putrescine (2+), spermidine (3+), spermine (4+) and cobalt hexamine (3+). The inter-DNA interaction potential is obtained with the umbrella sampling technique. The attractive force is rationalized in terms of the formation of ion bridges, *i.e.* multivalent ions which are simultaneously bound to the two opposing DNA molecules. The lifetime of the ion bridges is short on the order of a few nanoseconds.

Condensation of DNA induced by multivalent ions has been studied for many years [1]. Among the condensing agents, the polyamines constitute an important class. The polyamine-induced condensates of DNA (150 base pairs) have been shown to be liquid-crystalline with inter-helical spacing in the range 2.8-3.2 nm. Polyamines also induce the collapse of single DNA's into toroids [2]. Condensation has been reported for other polyelectrolytes as well, including actin and microtubules [3]. Although much work has been done to elucidate the mechanisms involved in stabilizing the condensed state, the detailed structural arrangement of the condensing agents is still unclear [4].

Multivalent ion-induced condensation cannot be explained with mean-field theory that always predicts a repulsive interaction between like-charged polyelectrolytes. Recent advances in the physics of strongly interacting systems go beyond the classical framework and it is now well established that dynamic correlation of cations shared by different polyanions gives rise to an attractive force [5] and the idea of a strongly correlated 2D liquid of adsorbed ions, similar to a Wigner crystal, has been proposed [6,7]. In theoretical modeling, DNA is usually treated as a uniformly charged cylinder, the counterions as point or spherical charges, and water as a continuous dielectric medium [8,9,10]. These approximations are appropriate for interactions over larger distances exceeding the atomic scale, but in dense systems, such as in DNA condensates, a molecular description is necessary for an understanding of the condensation phenomenon. This can now be achieved with the molecular dynamics (MD) computer simulation method [11,12].

Polyamines are associated with the compaction of DNA and play a role in the metabolism in eukaryotic cells [13]. Putrescine (Pu), spermidine (Sd), and spermine (Sm) are linear polyamines with two cationic nitrogen charges located at the terminal ends. Sd and Sm are tri- and tetravalent, respectively, with one or two more nitrogen charges along the contour. We investigated the interaction between two parallel double-stranded DNA duplexes with MD simulations and umbrella sampling [14]. To further investigate the effects of charge and ligand structure, we have also done simulations with trivalent cobalt hexamine (Co). The simulations show an attractive force, which can be understood in terms of the formation of transient ion bridges, *i.e.* counterions which are simultaneously and temporarily bound to the two opposing DNA's. To the best of our knowledge, this is the first validation of multivalent ion induced DNA attraction with an atomic model including a molecular description of the solvent water.

All simulations were for salt free systems using a rectangular cell, which contains one or two identical DNA decamers in the *B*-form of 2 nm outer diameter (see Fig. 1). A randomly selected sequence of 10 base-pairs (G5AAGAGGCTA3-C3TTCTCCGAT5) was chosen. The DNA charge is neutralized with 10 di, 7 tri or 5 tetravalent counterions (excess cationic charge was compensated with chloride). The 3' end of each strand is connected to the periodic image of the 5' end along the Z-axis (periodic boundary condition). This setup mimics an infinite array of parallel ordered DNA in fibers or liquid crystals. Note that the periodicity along the longitudinal axis matches the helical twist of the duplex with 10 base-pairs per turn. Furthermore, the connectivity of the decamers set by the boundary condition inhibits bending fluctuations with wave lengths exceeding the 3.4 nm longitudinal repeat distance of the simulation box. A snapshot of the transverse cross-section is shown in Fig. 1. The AMBER (v. 98) force field was used to model the DNA molecule, while partial charges, bond lengths, and bond angles of the counterions were derived employing the AMBER strategy [15]. The simulation box contains 4956 water molecules described with the simple point charge (SPC) model [16]. Electrostatic interactions were treated by the particle mesh Ewald method and the temperature was controlled around 300 K with Berendsen coupling. The GROMACS software [17] with a fixed box volume and a time step of 2 fs was used. Each MD run lasted more than 20 ns.

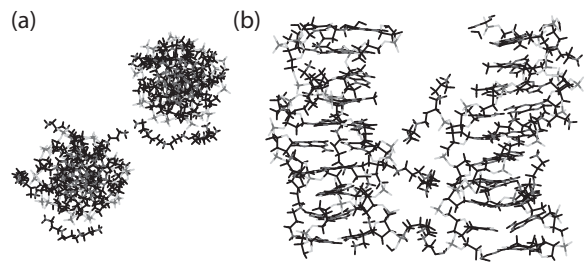

FIG. 1. (a): Top view of the simulation box with two parallel DNA decamers and ten Sm counterions in the initial configuration. The box has a transverse dimension of 7×7 nm$^2$ and 3.4 nm height. (b): Snapshot illustrating ion bridge formation.

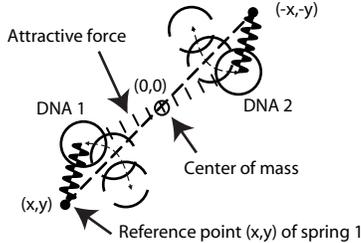

FIG. 2. Illustration of the cross section of the simulation box and how the two external springs pull the two DNA duplexes in opposite directions in the transverse plane.

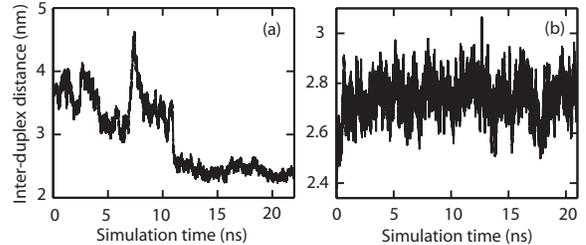

FIG. 3. Fluctuation in inter-duplex spacing of Sm-DNA. (a): simulation without springs; (b): as in (a) but with two springs centered at $(\pm x, \pm y) = (0.95, 0.95)$ nm.

DNA molecules form a side-by-side complex if they are attracting and unconstrained. In equilibrium the separation is small with the two duplexes almost touching each other. To study the interaction at larger separations, we have applied an external potential $P_{ext} = 1/2\, k\, (\Delta(x,y))^2$ with two springs. As shown in Fig. 2, these two springs pull the two duplexes in opposite directions. We have used a spring constant $k = 1000$ kJ mol$^{-1}$ nm$^{-2}$ and $\Delta(x,y)$ is the deviation of the pull group with respect to a reference point $(\pm x, \pm y)$. Since the two springs have the same spring constant, the total system experiences no net force. To obtain the interaction energy as a function of the distance $D_{int}$ between the centers of mass of the duplexes the contribution from $P_{ext}$ has to be subtracted from the interaction energy. In practice, we have assigned a weighing factor $\exp\left(P_{ext}/k_B T\right)$ to every sampling point. We then obtained the weighed probability distribution $\Omega_{weighed}(D_{int})$ from the fractional time the duplexes are separated by a distance $D_{int}$ and the true interaction energy follows from $F = -kT \ln \Omega_{weighed}$.

In order to study the DNA-counterion interaction without the influence of other DNA molecules, we have first done a simulation of a single DNA duplex with Sm counterions. The DNA molecule was positioned in the center of the box and the counterions were randomly distributed. In the first few nanoseconds, all Sm ions diffused towards the duplex and then remained *territorially* bound in the following few tens of nanoseconds with at least one of their four cationic nitrogen charges close to a phosphate moiety. In agreement with earlier results [18], the interaction was observed to be unspecific with territorial binding whereby the Sm's remain mobile and dynamic. A typical lifetime of a configuration in which a Sm ion is in close contact with the duplex is a few nanoseconds.

Next, we have done a simulation of two DNA duplexes and ten Sm counterions. The initial configuration was generated using the final state of the single DNA molecule simulation with all counterions territorially bound to DNA (see Fig. 1). The inter-helical distance was initially set to 3.8 nm. This distance does not allow a simultaneous contact of one Sm molecule with the two duplexes (the contour length of Sm is 1.6 nm). Due to the periodic boundary conditions and the fact that the top and bottom base-pairs of each DNA decamer are connected, the duplexes can hardly bend and they remain parallel.

The fluctuating inter-helical distance $D_{int}$ in a simulation of two DNA duplexes with Sm and without springs is displayed in Fig. 3. Initially, the two duplexes exhibited no correlated lateral motion. However, after 12 ns the duplexes formed a side-by-side complex and from then onwards they moved coherently with an inter-duplex separation of about 2.4 nm. Close inspection of the configurations revealed the details of the attraction (an example is displayed in Fig. 1). A Sm is usually territorially bound to one duplex. Since Sm is a linear tetravalent polyamine with a positive charge at each end, there is a dangling end jutting outwards in the surrounding medium. This dangling end can now be territorially bound to the other duplex and form an ion bridge. Note that the bridge is only temporarily formed; there is a continuous rearrangement of the bridging Sm. We surmise that the formation of these transient ion bridges results in a net attraction. Simulations were also done for Pu, Sd and Co. The result for the trivalent Sd is qualitatively similar. Control simulations with sodium counterions only, confirmed the absence of attraction and resulted in equilibrium separations of 5 nm.

To obtain sufficient sampling for larger separations it is necessary to apply the umbrella sampling technique. Con-tinuous potential curves are accordingly obtained and shown in Fig. 4. For the ligands of valence three or greater, *i.e.* for Co, Sd, and Sm-DNA, the potential exhibits a broad and pronounced minimum at 2.1, 2.3 and 2.4 nm, respectively (results of Co-DNA are not shown). The positions of the minima agree with the inter-duplex separation in the side-by-side complex obtained in the simulations without external forces and are related to the structure of the ligands. The depth of the potential takes the values -16 (Co), -9 (Sm), and -6 (Sd) $k_B T$. With increasing valence and smaller ligand size, the interaction potential becomes more attractive. For very short separations the potential is always repulsive due to electrostatic and hard-core interactions. For larger separations, beyond the minimum, the potential is attractive and monotonously increases until it levels off for $D_{\text{int}} > 3$ nm. The range of interaction is significantly shorter than half the length of the diagonal of the simulation box (5 nm), so that possible effects of the periodic boundary conditions are insignificant. Note that the multivalent ion mediated interaction energy is an order of magnitude larger than the value based on screened electrostatics and a helical distribution of adsorbed monovalent counterions [19].

The interaction in Pu-DNA is also attractive with a potential depth less than $2 k_B T$. This weak attraction is consistent with the experimental observation that Pu cannot induce condensation [1] and the experimentally observed weak DNA attraction in the presence of divalent magnesium [20]. In a simulation of two DNA's with sodium counterions we have checked that the potential is always repulsive. One should bear in mind that our simulations refer to salt-free systems with counterions only. We have checked that with the addition of monovalent salt (NaCl) the potential generally becomes less attractive and the minimum shifts to a larger inter-duplex distance. Furthermore, we have only considered a pair interaction. In a DNA condensate or liquid crystal one DNA molecule interacts with multiple DNA molecules and it is not *a priori* clear that the interactions are pair-wise additive [21]. For Sm-DNA in the absence of monovalent cations the experimental value of the inter-duplex distance is 2.8 nm [2]. This indicates the pair treatment of the interaction as a major cause for the shorter equilibrium inter-duplex distances as compared to the experimental values.

The inter-duplex force can be obtained from the derivative of the potential with respect to the separation. As an illustrative example, we have smoothed the data pertaining to Sm-DNA with the help of an arbitrary sixth order polynomial; the resulting force is shown in Fig. 4. The force can also be estimated in another way. The duplexes diffuse under the actions of the attractive force and the forces exerted by the springs. At the mean separation $\overline{D}_{\text{int}}$, these forces are balanced. If the springs are stretched by an amount $\Delta x$, the attractive force is approximately $\kappa \Delta x$. The inter-duplex direction is not al-ways co-linear with the directions of the springs but the deviations from co-linearity are always quite small and the resulting forces are consequently good first-order approximations. Good agreement with the curve as obtained from the derivative of the inter-duplex potential is obtained (see Fig. 4).

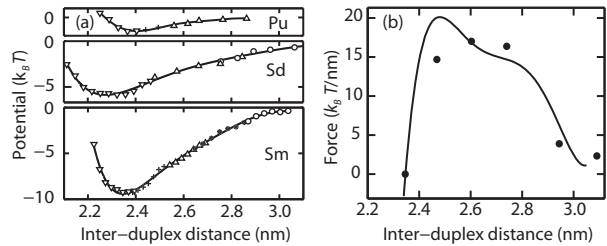

FIG. 4. (a): Interaction potential versus inter-duplex distance for Pu, Sd, and Sm-DNA. $\nabla$: no springs. With springs ($\pm x$, $\pm y$) = (0.90, 0.90), +; (0.95, 0.95), $\triangle$; (1.00, 1.00), $*$; (1.05, 1.05), $\bigcirc$ nm. (b): Force-separation curve for Sm-DNA. The dots are obtained from the spring extensions as described in the text.

As can already be gauged from the potential curve, for separations less than, say, 2.4 nm the inter-duplex force is repulsive. With increasing separation, the force becomes attractive, shows a maximum at around 2.5 nm, and eventually dwindles for distances larger than 3 nm. Note that the force is of short range on the order of the size of Sm (in a separate MD simulation the mean end-to-end distance of Sm was found to be 1.2 nm). For the trivalent ligands qualitatively similar results are obtained. These results comply with the notion that the attractive force is mediated through the formation of transient ion bridges between the interacting duplexes.

We now focus on the dynamics of the ligands and their spatial arrangement with respect to the duplexes to reveal the underlying molecular mechanism for the like-charge attraction. To characterize the dynamics we have monitored the time-dependence of the closest distance between *any* atom of the counterion and *any* atom of a duplex. The picture of ligand binding and dynamics in the simulations with two interacting duplexes is qualitatively the same as in the above described simulations with one DNA molecule. The minimum distance of closest approach is around 0.2 nm. We consider a ligand to be territorially bound to the duplex with one of its four charges close to a phosphate moiety if this minimum distance is less than 0.25 nm. The binding is highly dynamic; during the MD run, a counterion often attaches and detaches itself to the duplex and the lifetime of a bound configuration is typically no more than a few nanoseconds.

A ligand may be simultaneously bound to the two opposing duplexes, forming a temporary ion bridge. To establish a link between the attractive force and the formation of ion bridges, we have determined the number of bridging counterions by counting those which are within 0.25 nm from the two opposing duplexes at the same time. An example of the time-evolution of the total number of bridging Sm's is displayed in Fig. 5. Due to the continuous association and dissociation, the total number of bridges fluctuates about an average value $\bar{N}_{bridge}$ with a lifetime on the order of a few nanoseconds. Irrespective structure or charge, three or four ligands can be accommodated in the space between duplexes of 10 base-pairs length. At a particular inter-duplex distance the number of ion bridges does not depend on ligand charge and structure. The average number of bridges decreases with increasing mean inter-duplex separation. For separations larger than 3 nm, the size of the ligands becomes too small for the formation of an ion bridge.

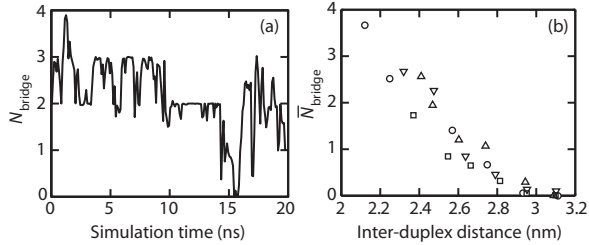

FIG. 5. (a): Time-evolution of the number of ion bridges in the Sm-DNA simulation with $(\pm x, \pm y) = (0.90, 0.90)$ nm (rebinned in 0.1 ns intervals). (b): Average number of ion bridges vs. inter-duplex distance.: ○, Co-DNA; △, Sm-DNA; ▽, Sd-DNA; □, Pu-DNA.

The range of the attractive force is understood in terms of ion bridges. Due to simultaneous binding to the counterions, the opposing and parallel duplexes are effectively connected and exert a force on each other. The depth of the potential is however determined by ligand charge and structure and a *simple* relationship with the number of bridges seems to be lacking. Furthermore, the ion bridges are transient with a short lifetime on the order of a few nanoseconds and characterized by a continuous rearrangement of the binding sites (*i.e.*, phosphate moieties on DNA and nitrogens of the polyamines). The typical lifetime of an ion bridge is hence of the same order of magnitude as the lifetime of a contact formed by a ligand and a single DNA molecule. Long range and persistent two-dimensional ordering of the associated counterions at the surface of the DNA molecule was not observed. In this respect our simulations do not support the notion of a strongly correlated 2D liquid of adsorbed ions.

Our simulated system differs from the situation in experimental studies in a number of aspects. First, the axes of interacting DNA duplexes are often skewed, because in a densely packed system the molecules undulate with a wavelength (deflection length) less than the persistence length [22]. We do not expect that the multivalent ion mediated interaction in a skewed configuration is qualitatively different from the one in a parallel configuration, because of the absence of long range position correlation among the adsorbed ions. Second, we only considered a pair interaction, whereas in a dense phase one DNA molecule interacts with multiple DNA molecules. Third, most experimental systems are not salt-free. Despite these obvious limitations, our simulations provide insight in the multivalent-ion induced attraction of DNA at the molecular level which is difficult, if not impossible, to obtain with alternative theoretical approaches or from experiments. The present study represents the first demonstration of the experimentally established counterion induced attraction using a full atomic model. This implies that this effect is now confirmed in a theoretical description beyond the dielectric continuum approximation of the solvent.

The support of research grants RG65/06 from Nanyang Technological University (to YM), T206B3207 from the MOE (to LN) and R144000145 from National University of Singapore (to JvdM) is acknowledged.